\documentclass[a4paper,12pt, epsfig]{article}
\usepackage{epsfig}
\usepackage{amssymb}
\usepackage{amsfonts}
\usepackage{amsmath}

\newskip\humongous \humongous=0pt plus 1000pt minus 1000pt

\newif\ifdtup

\catcode`\@=11

\@addtoreset{equation}{section}
\def\theequation{\thesection.\arabic{equation}}

\def\@normalsize{\@setsize\normalsize{15pt}\xiipt\@xiipt
\abovedisplayskip 14pt plus3pt minus3pt%
\belowdisplayskip \abovedisplayskip
\abovedisplayshortskip \z@ plus3pt%
\belowdisplayshortskip 7pt plus3.5pt minus0pt}

\def\small{\@setsize\small{13.6pt}\xipt\@xipt
\abovedisplayskip 13pt plus3pt minus3pt%
\belowdisplayskip \abovedisplayskip
\abovedisplayshortskip \z@ plus3pt%
\belowdisplayshortskip 7pt plus3.5pt minus0pt
\def\@listi{\parsep 4.5pt plus 2pt minus 1pt
     \itemsep \parsep
     \topsep 9pt plus 3pt minus 3pt}}

\relax

\catcode`@=12

\catcode`\@=11

\def\section{\@startsection{section}{1}{\z@}{3.5ex plus 1ex minus
   .2ex}{2.3ex plus .2ex}{\large\bf}}

\def\thesection{\arabic{section}}
\def\thesubsection{\arabic{section}.\arabic{subsection}}

\def\appendix{\setcounter{section}{0}
 \def\thesection{Appendix \Alph{section}}
 \def\thesubsection{\Alph{section}.\arabic{subsection}}
 \def\theequation{\Alph{section}.\arabic{equation}}}
\def\SymBoxes#1#2#3#4{\newdimen\un@t \un@t#3%
\raisebox{#1}{\rule{#2\un@t}{#4}\hskip-#2\un@t% lower horizontal
\@tempdimb\un@t \advance\@tempdimb by-#4\@tempcntb#2\relax%
\@whilenum{\@tempcntb>0}\do{%                         % #2 vertical lines
\rule{#4}{\un@t}\hskip\@tempdimb \advance\@tempcntb by\m@ne}%
\hskip-#2\un@t \rule[\un@t]{#2\un@t}{#4}%
\rule[\un@t]{#4}{#4}\hskip-#4%             % upper horizontal line
\rule{#4}{\un@t}}\hskip-#4}                % rightest vertical line
\begin{document}
%\begin{letter}{~}

%%%%%%Define some new commands and  macros
\newcommand{\beq}{\begin{equation}}
\newcommand{\eeq}{\end{equation}}
\newcommand{\bea}{\begin{eqnarray}}
\newcommand{\eea}{\end{eqnarray}}
\newcommand{\beas}{\begin{eqnarray*}}
\newcommand{\eeas}{\end{eqnarray*}}
\newcommand{\defi}{\stackrel{\rm def}{=}}
\newcommand{\non}{\nonumber}
\newcommand{\bquo}{\begin{quote}}
\newcommand{\enqu}{\end{quote}}
%%%%%%%%%%%%%%%%
\renewcommand{\(}{\begin{equation}}
\renewcommand{\)}{\end{equation}}
%%%%%%%%%%%%%%%%%%%%%%%%%%%%%%%%%% definitions
\def\IZ{{\mathbb Z}}
\def\IR{{\mathbb R}}
\def\IC{{\mathbb C}}
\def\IQ{{\mathbb Q}}
\def\Rhat{{\hat R}}
\def\Chat{{\hat C}}

\def \eqn#1#2{\begin{equation}#2\label{#1}\end{equation}}
\def\de{\partial}
\def\Tr{ \hbox{\rm Tr}}
\def\H{ \hbox{\rm H}}
\def\HE{ \hbox{$\rm H^{even}$}}
\def\HO{ \hbox{$\rm H^{odd}$}}
\def\K{ \hbox{\rm K}}
\def\Im{ \hbox{\rm Im}}
\def\Ker{ \hbox{\rm Ker}}
\def\const{\hbox {\rm const.}}
\def\o{\over}
\def\im{\hbox{\rm Im}}
\def\re{\hbox{\rm Re}}
\def\bra{\langle}\def\ket{\rangle}
\def\Arg{\hbox {\rm Arg}}
\def\Re{\hbox {\rm Re}}
\def\Im{\hbox {\rm Im}}
\def\exo{\hbox {\rm exp}}
\def\diag{\hbox{\rm diag}}
\def\longvert{{\rule[-2mm]{0.1mm}{7mm}}\,}
\def\a{\alpha}
\def\dag{{}^{\dagger}}
\def\tq{{\widetilde q}}
\def\p{{}^{\prime}}
\def\W{W}
\def\N{{\cal N}}
\def\hsp{,\hspace{.7cm}}
\newcommand{\C}{\ensuremath{\mathbb C}}
\newcommand{\Z}{\ensuremath{\mathbb Z}}
\newcommand{\R}{\ensuremath{\mathbb R}}
\newcommand{\rp}{\ensuremath{\mathbb {RP}}}
\newcommand{\cp}{\ensuremath{\mathbb {CP}}}
\newcommand{\vac}{\ensuremath{|0\rangle}}
\newcommand{\vact}{\ensuremath{|00\rangle}}
\newcommand{\oc}{\ensuremath{\overline{c}}}
\begin{titlepage}
\begin{flushright}
%ULB-TH/mm-dd\\
%hep-th/yymmnnn\\
\end{flushright}
\bigskip
\def\thefootnote{\fnsymbol{footnote}}

\begin{center}
{\Large {\bf
Multiparametric Quantum Algebras \\}}
\vspace{0.1cm}
{\Large {\bf and the Cosmological Constant}}\\
\end{center}

\bigskip
\begin{center}
{\large  Chethan
KRISHNAN$^{1}$\footnote{\texttt{Chethan.Krishnan@ulb.ac.be}}
and Edoardo DI NAPOLI$^{2}$
\footnote{\texttt{edodin@physics.unc.edu}}}\\
\end{center}

\renewcommand{\thefootnote}{\arabic{footnote}}

\begin{center}
\vspace{1em}
{\em  $^{1}${ International Solvay Institutes,\\
Physique Th\'eorique et Math\'ematique,\\
ULB C.P. 231, Universit\'e Libre
de Bruxelles, \\ B-1050, Bruxelles, Belgium\\}}
\vspace{1em}
{\em $^{2}${ Department of Physics and Astronomy \\
CB\# 3255 Phillips Hall \\
University of North Carolina \\
Chapel Hill, NC 27599-3255, USA \\}}

\end{center}

\noindent
\begin{center} {\bf Abstract} \end{center}
With a view towards applications for de Sitter, we construct the 
multi-parametric $q$-deformation of the $so(5,\IC)$ algebra using the 
Faddeev-Reshetikhin-Takhtadzhyan(FRT) formalism.
%If one tries to view de Sitter as a true (as opposed to a meta-stable)
%vacuum, there is a tension between the finiteness of its entropy and the
%infinite-dimensionality of its Hilbert space. We invetsigate the 
%viability
%of one proposal to reconcile this tension using $q$-deformation. After
%defining a differential geometry on the quantum de Sitter space, we try 
%to
%constrain the value of the deformation parameter by imposing the 
%condition
%that in the undeformed limit, we want the real form of the (inherently
%complex) quantum group to reduce to the usual SO(4,1) of de Sitter. We
%find that this forces $q$ to be a real number. Since it is known that
%quantum groups have finite-dimensional representations only for $q=$ root
%of unity, this suggests that standard $q$-deformations cannot give rise 
%to
%finite dimensional Hilbert spaces, ruling out finite entropy for
%$q$-deformed de Sitter.

%\begin{center}
\vspace{1.6 cm}

\vfill

\end{titlepage}
\bigskip

\hfill{}
\bigskip

\tableofcontents

\setcounter{footnote}{0}

%\keywords{Quantum groups, Hopf algebras, de Sitter space}

%\received{???????? ?st, 2000} \accepted{???????? ?th, 1998}
%\preprint{UTTG-01-06}

\section{\bf Introduction}

There are reasons based on arguments of holography and finiteness of entropy, to believe that 
the Hilbert space for quantum theory in a de Sitter background is finite 
dimensional 
\cite{Willy, Banks, Bousso}. Since the 
isometry group of de Sitter, $SO(4,1)$, has to be represented on this 
Hilbert space, and since 
we expect that quantum theory is unitary, this gives rise to an immediate problem: $SO(4,1)$ 
cannot have finite-dimensional unitary representations, because it is a non-compact group. 
It is in 
this context that the possibility of considering a deformed de Sitter 
space with a 
$q$-deformed version 
of its isometry group, becomes interesting \cite{Pouliot, Lowe1, 
Guijosa:2005qi, Lowe2} (Some of these references work in the context of 
dS/CFT). It is a well-known fact that for 
certain values of the 
deformation parameter, (non-compact) quantum groups have unitary, finite 
dimensional 
representations \cite{Stein1, Stein2, Stein3}.

But recently it was shown \cite{Krishnan:2006bq} that single-parameter 
quantum deformation can give rise to deformed de Sitter space only when 
the 
deformation 
parameter is real. This throws a spanner in the above program because 
finite dimensional represenations for one-parameter deformations exist 
only when the deformation parameter $q$ is a root of unity. One obvious 
way to work around this problem is to consider multi-parametric 
deformations of the de Sitter isometry group, and the aim of this paper is 
to take a first step in that direction by writing down the algebra for 
this case explicitly.

Another reason why multi-parametric deformations are interesting is 
because in the coordinate system 
of a static observer in de Sitter, the full $SO(4,1)$ isometry group 
is not visible: the manifest isometries are $SO(3)$ 
and a time-translation (see the Appendix for an elemantary demonstration 
of this fact). So one of the questions we need 
to answer when we quantize in de Sitter, is to understand how the static 
observer and the full isometry group are related to each other. One hope 
behind the construction of multi-parametric deformations of $SO(4,1)$, 
is that finding representations of such an algebra will shed some light on
the states of the observer and their relation to the representations of the 
full isometry group. We will 
be working 
at the level of complexified algebras, so what we refer to as the algebra of $SO(4,1)$ or 
$SO(5)$ is in fact $B_2$ in the Cartan scheme.

The usual one-parameter $q$-deformation for a Lie Algebra is the  
Drinfeld-Jimbo (DJ) Algebra. We 
will be interested in a construction of this algebra starting with a dual 
description in terms of $R$-matrices:  
using the Faddeev-Reshetikhin-Takhtadzhyan \cite{FRT} approach. What we will do in this paper is to take 
the DJ 
algebras to be defined by the FRT method, and then extend the definition by using a 
generalized, multi-parametric $R$-matrix \cite{Reshet, Schirr}. We will do this explicitly 
for 
$SO(5,\IC)$ and the result will be a multi-parametric generalization of 
the DJ algebra. %for $B_2$.

In the next section we will provide an introduction to the DJ algebra and how 
it can be derived from a dual description. In section 4, we will write down the explicit form 
of the multi-parametric R-matrix for $SO(5)$ from \cite{Schirr}, and use that in the dual 
description to construct the multi-parametric algebra for $SO(5)$. We 
conclude with some speculations and possibilities for future 
research.

Finite-dimensionality of de Sitter Hilbert space has also been discussed 
in
\cite{Parikh:2004ux, Parikh:2004wh}, and $q$-deformation in the context of
AdS/CFT has been considered in \cite{Jevicki:2000ty, Corley:2002ny}.

\section{\bf One-parameter DJ Algebra and its Dual Description}

Drinfeld-Jimbo algebra is a deformation of the universal enveloping algebra of the Lie 
algebra of a classical group. A universal enveloping algebra is the algebra spanned 
by polynomials in the generators, modulo the commutation relations. When we deform it, 
we mod out by a set of deformed relations, instead of the usual commutation relations. 
These relations are what define the DJ algebra. When the deformation parameter tends to 
the limit unity, the algebra reduces to the universal enveloping algebra of the usual Lie 
algebra. 

We will write down the algebra relations in the so-called Chevalley-Cartan-Weyl basis. The rest of the 
generators of the Lie Algebra can be generated through commutations between 
these. The Drinfeld-Jimbo algebra is constructed as a deformation of the relations between 
the Chevalley generators. So without any further ado, lets write down the form 
of the DJ algebra \cite{Klimyk} for a generic semi-simple Lie algebra ${\bf g}$ of rank $l$ and Cartan matrix 
($a_{ij}$). In what follows, $q$ is a fixed non-zero complex number (the deformation 
parameter), and $q_i=q^{d_i}$, 
with $d_i=(\alpha_i,\alpha_i)$ where $\alpha_i$ are the simple roots of the Lie algebra. 
The norm used in the definition of $d_i$ is the norm defined in the dual space of the 
Cartan sub-algebra, through the Killing form. These are all defined in standard 
references \cite{cahn, Klimyk}. The indices run from $1$ to $l$. 

With these at hand we can define the Drinfeld-Jimbo 
algebra $U_q({\bf g})$ as the algebra generated by $E_i$, $F_i$, $K_i$, $K_i^{-1}$, $1\le i \le l$, and 
the defining relations,
\eqn{DJ1}{K_iK_j=K_jK_i, \  K_iK_i^{-1}=K_i^{-1}K_i, }
\eqn{DJ2}{K_iE_j={q_i}^{a_{ij}}E_jK_i, \ K_iF_j={q_i}^{-a_{ij}}F_jK_i,}
\eqn{DJ3}{E_iF_j-F_jE_i=\delta_{ij}\frac{K_i-{K_i}^{-1}}{q_i-q_i^{-1}},}
\eqn{DJSerre1}{\sum_{r=0}^{1-a_{ij}}(-1)^r [[1-a_{ij}; 
r]]_{q_i}E_i^{1-a_{ij}-r}E_jE_i^r=0,  \  i\neq j,}
\eqn{DJSerre2} {\sum_{r=0}^{1-a_{ij}}(-1)^r [[1-a_{ij};
r]]_{q_i}F_i^{1-a_{ij}-r}F_jF_i^r=0,  \  i\neq j,}
with 
\eqn{auxiliary1}{\big[\big[n;r\big]\big]_q=\frac{[n]_q!}{[r]_q![n-r]_q!}}
and
\eqn{auxiliary2}{[n]_q=\frac{q^n-q^{-n}}{q-q^{-1}}, \ [n]_q!=[1]_q[2]_q...[n]_q, \ 
[0]_q\equiv 1.}

The relations containing only the $E$s or the $F$s are called Serre relations and 
they should be thought of as the 
price that we have to pay in order to write the algebra relations entirely in terms of the 
Chevalley generators. Sometimes, it is useful to write $K_i$ as $q_i^{H_i}$. In the limit 
of $q\rightarrow 1$, the DJ algebra relations reduce to the Lie Algebra relations 
written in the Chevalley basis, with $H_i$'s the generators in the Cartan 
subalgebra and the $E_i$'s and $F_i$'s the raising and lowering operators.   

We will be interested in the specific case of $SO(5)$ (Cartan's $B_2$), and we will 
rewrite the DJ 
algebra $U_{q^{1/2}}({\bf so(5)})$ for that case in a slightly different form for later convenience:
\eqn{so51}{k_1k_2=k_2k_1, \ k_1^{-1}=q^{H_1+H_2/2}, \ k_2^{-1}=q^{H_2/2}}
\eqn{so52}{k_1E_1=q^{-1}E_1k_1, \ k_2E_1=qE_1k_2,}
\eqn{so53}{k_1E_2=E_2k_1, \ k_2E_2=q^{-1}E_2k_2,}
\eqn{so54}{k_1F_1=qF_1k_1, \ k_2F_1=q^{-1}F_1k_2,}
\eqn{so55}{k_1F_2=F_2k_1, \ k_2F_2=qF_2k_2,}
\eqn{so56}{[E_1,F_1]=\frac{k_2k_1^{-1}-k_2^{-1}k_1}{q-q^{-1}}, \ 
[E_2,F_2]=\frac{k_2^{-1}-k_2}{q^{1/2}-q^{-1/2}}.}
The Serre relations take the form:
\eqn{so5s1}{E_1^2E_2-(q+q^{-1})E_1E_2E_1+E_2E_1^{2}=0}
\eqn{so5s2}{E_1E_2^{3}-(q+q^{-1}+1)E_2E_1E_2^2+(q+q^{-1}+1)E_2^2E_1E_2-E_2^3E_1=0}
with analogous expressions for the $F$s.

Drinfeld-Jimbo algebra is one way to describe a ``quantum group". Another way to do 
this is to work with the groups directly and deform the group structure using the 
so-called $R$-matrices, rather than to deform the universal envelope of the Lie 
algebra. It turns out that both these approaches are dual to each other, and one can 
obtain the DJ algebra by starting with $R$-matrices. Faddeev-Reshetikhin-Takhtadzhyan 
have constructed a formalism for working with the $R$-matrices, and to construct the 
DJ algebra starting from the dual approach. So, a natural place to look for when 
trying to generalize the DJ algebra of $SO(5)$ is to look at this dual construction 
and try to see whether it admits any generalizations. 

In the rest of this section, we will review the construction of the DJ algebra 
starting with the $R$-matrices. In the next section, we will start with a 
multi-parametric generalization of the $R$-matrix for $SO(5)$ and follow an analogous 
procedure to obtain the multi-parametric $SO(5)$ DJ algebra.

As already mentioned, the deformation of the group structure is done in the dual 
picture through the introduction of the $R$-matrix. The duality between the two 
approaches is manifested through the so-called $L$-functionals \cite{Klimyk}. If one 
defines the $L$-functionals as certain matrices constructed from the DJ algebra 
generators, then the $R$-matrix and the $L$-functionals would together satisfy 
certain relations (which we will call the duality relations), as a consequence of the 
fact that the generators satisfy the DJ algebra. 
Conversely, we could start 
with $L$-functionals thought of as matrices with previously unconstrained matrix 
elements, and then the duality relations would be the statement that the matrix 
elements should satisfy 
the DJ algebra. Thus, the $L$-functionals, together with the duality relations is 
equivalent to the DJ algebra.

For any $R$-matrix\footnote{It is useful here to keep in mind that 
$R$-matrices are $N^2\times N^2$ matrices.}, we can define an algebra 
$A(R)$, with $N(N+1)$
generators $l^{+}_{ij}$, $l^{-}_{ij}$, $i\le j, j=1,2,...,N$, and defining
relations
\eqn{gen1}{L^{\pm}_{1}L^{\pm}_{2}R=RL^{\pm}_{2}L^{\pm}_{1}, \
L^{-}_1L^{+}_2R=RL^{+}_2L^{-}_1 }
\eqn{gen2}{l^{+}_{ii}l^{-}_{ii}=l^{-}_{ii}l^{+}_{ii}=1, \ i=1,2,...,N}
where the matrices $L^{\pm}\equiv (l^{\pm}_{ij})$ and $l^{+}_{ij}=0=l^{-}_{ji}$, for $i > j$
(that is, they are upper or lower triangular).
The subscripts 1 and 2 have the following meaning:
$L_1^+$
stands for $L^+$ tensored with the $N \times N$ identity matrix, and $L_2^+$ stands
for the $N \times N$ identity matrix tensored with $L^+$. So the matrix
multiplication with $R$ is well-defined because
the $R$-matrix is an $N^2 \times N^2$ matrix. The above relations will be referred to as 
the duality relations.
It turns out that this algebra has a Hopf algebra structure with 
\begin{eqnarray}
{\rm Comultiplication:} \ &&
\Delta (l^{\pm}_{ij})=\sum_k l^{\pm}_{ik} \otimes l^{\pm}_{kj}, 
\label{1st}\\
{\rm Counit:} \ && \epsilon(l^{\pm}_{ij})=\delta_{ij}, \\ 
{\rm Antipode:} \ && S(L^{\pm})=(L^{\pm})^{-1}. \label{3rd}
\end{eqnarray}

Now, lets choose the $R$-matrix in the above case to be the one-parameter
$R$-matrix for $SO(N)$, with $N=2n+1$. 
\begin{eqnarray}
\nonumber R&=&q\sum_{i\neq i^{\prime}}^{2n}E_{ii}\otimes
E_{ii}+q^{-1}\sum_{i\neq i^{\prime}}^{2n}E_{ii}\otimes E_{i^{\prime}i^{\prime}}
+E_{n+1,n+1}\otimes E_{n+1,n+1}+ \\
&+&\sum_{i \neq j, j^{\prime}}^{2n}E_{ii}\otimes E_{jj} + (q-q^{-1})\Big[ \sum_{i>j}^{2n}E_{ij}\otimes E_{ji} -
\sum_{i>j}^{2n}q^{\rho_i-\rho_j}E_{ij}\otimes E_{i^{\prime}j^{\prime}} \Big].
\end{eqnarray}
Here $E_{ij}$ is the $2n \times 2n$ matrix with 1 in the $(i,j)$-position and 0
everywhere else, and the symbol $\otimes$ stands for tensoring of two matrices.
$i^{\prime}=2n+2-i$ and similarly for $j^{\prime}$. The deformation parameter is $q$. Finally, 
$(\rho_1,\rho_2,...,\rho_{2n})=(n-1/2,
n-3/2,...,1/2,0,-1/2,...,-n+1/2)$.

Let $I(so(N))$ be the two-sided ideal in
$A(R)$
generated by
\eqn{ideal1}{L^{\pm}C^{t}(L^{\pm})^t(C^{-1})^t=I=C^t(L^{\pm})^t(C^{-1})^tL^{\pm}}
where $I$ is the identity matrix, and the metric $C$ defines a length in the
vector space where the quantum matrices are acting. $C$ provides the constraint
arising from the fact that the underlying classical group is an orthogonal group:
$TC^{-1}T^{t}C=I=C^{-1}T^tCT$ for quantum matrices $T$ (see \cite{Schirr}). For
$SO(N)$,
\eqn{ortho1}{C=(C^i_j), \ C^i_j=\delta_{ij^{\prime}}q^{-{\rho}_i}}
with $j^{\prime}$ and ${\rho}_i$ are as defined above.

Now, $I(so(N))$ is a Hopf ideal of $A(R)$ \cite{Klimyk}, so the quotient $A(R)/I(so(N))$ is also
a Hopf algebra which we will call $U_q^{L}({\bf so(N)})$. Now, there is a 
theorem (see, for example \cite{Klimyk} or \cite{FRT} for a proof) which
says that $U_q^{L}({\bf so(N)})$ is isomorphic to $U_{q^{1/2}}({\bf so(2n+1)})$, which is the DJ
algebra for $SO(2n+1)$ with deformation parameter $q^{1/2}$. Explicitly, this isomorphism can be 
written down as, 
\begin{eqnarray}
l^{+}_{ii}=q^{-H^{\prime}_{i}}, \ l^{+}_{i^{\prime}i^{\prime}}=q^{H^{\prime}_{i}}&,& \ 
l^{+}_{n+1,n+1}=l^{-}_{n+1,n+1}=1, \\
l^{+}_{k,k+1}=(q-q^{-1})q^{-H^{\prime}_k} E_k &,& \ l^{+}_{2n-k+1,2n-k+2}=-(q-q^{-1})
q^{H^{\prime}_{k+1}}E_k, \\
l^{-}_{k+1,k}=-(q-q^{-1}) F_k q^{H^{\prime}_{k}} &,& \ l^{-}_{2n-k+2,2n-k+1}=(q-q^{-1}) F_k q^{-H^{\prime}_{k+1}}, \\
l^{+}_{n,n+1}&=&(q^{1/2}+q^{-1/2})^{1/2}(q^{1/2}-q^{-1/2})q^{-H^{\prime}_{n}}E_n,  \\ 
l^{+}_{n+1,n+2}&=&-q^{-1/2}(q^{1/2}+q^{-1/2})^{1/2}(q^{1/2}-q^{-1/2})E_n , \\
l^{-}_{n+1,n}&=&-(q^{1/2}+q^{-1/2})^{1/2}(q^{1/2}-q^{-1/2})F_n q^{H^{\prime}_{n}}, \\
l^{-}_{n+2,n+1}&=&q^{1/2}(q^{1/2}+q^{-1/2})^{1/2}(q^{1/2}-q^{-1/2})F_n 
\end{eqnarray}
Here, $i=1,2,...,n$ as always, and $1 \le k \le n-1$. $H^{\prime}_i=H_i+H_{i+1}+...
+H_{n-1}+H_n/2$. The above relations (which we will call the isomorphism relations) define the relations between elements of the $L$ 
matrices and the Chevalley-Cartan-Weyl generators. Sometimes it will be convenient to call 
$q^{-H^{\prime}_i}$ as $k_i$ because it makes comparison with $SO(5)$ DJ algebra (written 
earlier) more direct. 

\section{\bf The Multi-parametric Algebra}

Our procedure for constructing the multi-parametric algebra is 
straightforward. Instead of using the usual one-parametric $R$-matrices in 
the duality relations, we use the multi-parametric $R$-matrices that 
Schirrmacher has written down \cite{Schirr}. We keep the isomorphism relations to be 
the same as above and use the duality relations to define the new multi-parametric 
algebra.
 
In principle this procedure could be done 
for all the multi-parametric $R$-matrices of all the different Cartan groups using 
their associated isomorphism relations. We have 
endeavored to do this procedure for only the case of $SO(5)$, but at least for the 
smaller Cartan groups, the exact same procedure can be performed on a computer using the appropriate 
$R$-matrices.  
To write down the form of the multi-parametric DJ algebra for a generic semisimple Lie algebra is an interesting 
problem which we have not attempted to tackle here.

The multi-parametric $R$-matrix for $SO(2n+1)$ (which for our purposes is the 
same thing as $B_n$) looks 
like
\begin{eqnarray}
\nonumber R&=&r\sum_{i\neq i^{\prime}}^{2n}E_{ii}\otimes 
E_{ii}+r^{-1}\sum_{i\neq i^{\prime}}^{2n}E_{ii}\otimes E_{i^{\prime}i^{\prime}}
+E_{n+1,n+1}\otimes E_{n+1,n+1}+ \\
&+&\sum_{i < j, \ i \neq 
j^{\prime}}^{2n}\frac{r}{q_{ij}}E_{ii}\otimes E_{jj} + \sum_{i > j, \ i \neq
j^{\prime}}^{2n}\frac{q_{ij}}{r}E_{ii}\otimes E_{jj} + \nonumber \\
&+&(r-r^{-1})\Big[ \sum_{i>j}^{2n}E_{ij}\otimes E_{ji} - 
\sum_{i>j}^{2n}r^{\rho_i-\rho_j}E_{ij}\otimes E_{i^{\prime}j^{\prime}} \Big].
\end{eqnarray}
The deformation parameters 
are $r$ and $q_{ij}$ and they are not all 
independent: $q_{ii}=1$, $q_{ji}=r^2/q_{ij}$ 
and $q_{ij}=r^2/q_{ij^{\prime}}=r^2/q_{i^{\prime}j}=q_{i^{\prime}j^{\prime}}$. 
These relations basically imply that $q_{ij}$ with $i<j\leq n$ determine all the 
deformation parameters. It should be noted that when all the independent deformation parameters 
are set 
equal to each other ($=q$), then the $R$-matrix reduces to the usual one 
parametric version. In the case of $SO(5)$, the multi-parametric $R$-matrix has only 
two independent parameters, which we will call $r$ and $q$.

We extensively used a Mathematica package called NCALGEBRA (version 3.7)\cite{NC} to do the 
computations, 
since the matrix elements (being generators of an algebra) are not commuting objects. The 
first task is to obtain the duality relations between the matrix elements explicitly. 
The $L$ matrices are chosen to be upper and lower triangular. The task is straightforward 
but tedious because the duality relations are 25 by 25 matrix relations for
the case of $SO(5)$. So one has to scan through the resulting output to 
filter out the relations that are dual to the relations between the Chevalley-Cartan-Weyl 
generators. Doing the calculation for the single-parameter case will give a hint about 
which relations are relevant in writing down the algebra. 

The first line of the isomorphism 
relations (for the specific case of $SO(5)$) implies that we 
can use $k_1, k_2, 1, k_2^{-1}$ and $k_1^{-1}$ instead of $l_{11}, l_{22}, l_{33}, l_{44}$ 
and $l_{55}$ respectively. With this caveat, the algebra looks like the following in terms of the 
relevant $L$ matrix elements:
\begin{eqnarray}
k_1k_2&=&k_2k_1, \\
k_1l^{+}_{12}=\frac{r}{q^2}l^{+}_{12}k_1 &,& k_2l^{+}_{12}=\frac{q^2}{r}l^{+}_{12}k_2, \\
k_1l^{+}_{23}=\frac{q}{r}l^{+}_{23}k_1 &,& k_2l^{+}_{23}=q^{-1}l^{+}_{23}k_2, \\
k_1l^{-}_{21}=rl^{-}_{21}k_1 &,& k_2l^{-}_{21}=\frac{r}{q^2}l^{-}_{21}k_2, \\
k_1l^{-}_{32}=\frac{q}{r}l^{-}_{32}k_1 &,&  k_2l^{-}_{32}=ql^{-}_{32}k_2, \\
\left[l^{+}_{45},l^{-}_{21}\right]&=&(q-q^{-1})(k_1^{-2}-k_2^{-2}), \\ 
\left[l^{+}_{23},l^{-}_{32}\right]&=&(q-q^{-1})(k_2-k_2^{-1}), \\
{l^{+}_{12}}^2l^{+}_{23}-\Big(\frac{q}{r}&+&\frac{r}{q^3}\Big)l^{+}_{12} l^{+}_{23} l^{+}_{12}
+\frac{1}{q^2}l^{+}_{23}{l^{+}_{12}}^2=0, \\
\frac{q^2}{r^2}{l^{+}_{23}}^3-\Big(\frac{q^2}{r}+\frac{q^5}{r^3}+\frac{q}{r}\Big)
{l^{+}_{23}}^2l^{+}_{12}l^{+}_{23}&+&\Big(q+\frac{q^4}{r^2}+\frac{q^5}{r^2}\Big)l^{+}_{23}
l^{+}_{12}{l^{+}_{23}}^2-\frac{q^4}{r}l^{+}_{12}{l^{+}_{23}}^3=0.
\end{eqnarray}
The last two equations correspond to the Serre relations. (We write them down only for the 
$L^{+}$ matrix elements.). As an example of the general procedure for obtaining these algebra 
relations from the duality relations (i.e., the Mathematica output), we will demonstrate 
the derivation of the first Serre relation. The 
relevant expressions that one gets from Mathematica are:
\begin{eqnarray}
l^{+}_{12}l^{+}_{23}-\frac{q}{r}l^{+}_{23}l^{+}_{12}&=&-(q-q^{-1})l^{+}_{13}k_2, \\
l^{+}_{12}l^{+}_{13}&=&\frac{1}{q}l^{+}_{13}l^{+}_{12}. 
\end{eqnarray}
Solving for $l^{+}_{13}$ from the first equation by multiplying by $k_2^{-1}$ on the 
right, plugging it back into the second equation and using the commutation rules for $k_2$,
we get our Serre relation. This kind of manipulation is fairly typical in the derivation 
of the above algebra. 

As a next step, we use the isomorphism relations defined at the end of the last section to 
rewrite the above algebra in terms of the Chevalley-Cartan-Weyl type generators. The result is
\begin{eqnarray}
k_1k_2&=&k_2k_1, \\
k_1E_1=\frac{r}{q^2} E_1k_1 &,& k_2E_1=\frac{q^2}{r}E_1k_2, \\
k_1E_2=\frac{q}{r}E_2k_1 &,& k_2E_2=\frac{1}{q}E_2k_2, \\
k_1F_1=rF_1k_1 &,& k_2F_1=\frac{r}{q^2}F_1k_2, \\
k_1F_2=\frac{q}{r}F_2k_1 &,& k_2F_2=qF_2k_2, \\
\frac{q}{r}E_1F_1-\frac{r}{q}F_1E_1&=&\frac{k_2k_1^{-1}-k_2^{-1}k_1}{q-q^{-1}}, \\
\left[E_2,F_2 \right]&=&\frac{k_2^{-1}-k_2}{q^{1/2}-q^{-1/2}}, \\
E_1^2E_2-\Big(\frac{q^2}{r}&+&\frac{r}{q^2}\Big)E_1E_2E_1+E_2E_1^2=0, \\
E_2^3E_1-\Big(\frac{r}{q}+\frac{q^2}{r}+\frac{r}{q^2}\Big)E_2^2E_1E_2&+&\Big(
\frac{r^2}{q^3}+1+q\Big)E_2E_2E_2^2-\frac{r}{q}E_1E_2^3=0. 
\end{eqnarray}
This, is our final form for the multi-parametric version of $SO(5)$ 
Drinfeld-Jimbo algebra. Together with the Hopf algebra relations from 
(\ref{1st})-(\ref{3rd}), these relations completet our definition of the 
multi-pramateric 
algebra. 
Notice that they reduce to the one-parameter DJ algebra of $SO(5)$ in the 
limit of 
$r \rightarrow q$.

\section{\bf Results and Outlook}
We have constructed the multi-parametric version of the Drinfeld-Jimbo algebra for 
the case of $SO(5)$ with the intention of possible applications in 
de Sitter quantum 
mechanics and quantum gravity. As physicists, we are more interested in 
working with the algebra 
directly than working with the groups and the $R$-matrix, because presumably, 
finding representations of the 
algebra would be more direct (even though still non-trivial). Finding 
representations 
is interesting because 
that could be a first step 
in 
embedding the Hilbert space of the static patch 
of an observer, in the Hilbert space of the full de Sitter space. 
It might be the 
case that embedding the $SO(3)_q$ of the observer is easier to 
accomplish, in the 
added luxury of two parameters. Also, if it turns out that this embedding 
is 
possible only when there is a relationship between the parameters, it 
could 
translate into a statement about the surprising smallness of the 
cosmological 
statement in terms of scales which are more readily accessible to the 
observer. Of 
course, at this stage, this is pure speculation. The bottomline is that it 
seems like there is the exciting possibility of addressing the 
problem of the smallness of the positive cosmological 
constant using the multi-parametric deformation\footnote{We thank Willy 
Fischler for suggesting this to us.}. Some of these issues are 
currently being investigated.

It is also interesting as a pure mathematical problem to write down the 
multi-parametric DJ 
algebra for a generic Lie algebra. To the best of our knowledge, this is 
still an 
open 
problem.
 
\section{Acknowledgments}

This project started out and evolved through conversations with Willy Fischler. Its 
a pleasure to thank him for his help, and his willingness to share ideas, 
including the suggestion to look at multi-parametric quantum groups. We would like 
to thank Hyuk-Jae Park and Marija Zanic for stimulating conversations and Uday 
Varadarajan for suggesting useful references. R. Jaganathan was kind 
enough to mail us the vital and hard-to-get 
reference \cite{FRT}. 
This material is based upon work supported by the National Science Foundation under
Grant Nos. PHY-0071512 and PHY-0455649, and with grant support from the US Navy, Office
of Naval Research, Grant Nos. N00014-03-1-0639 and N00014-04-1-0336, Quantum Optics
Initiative.

\section{\bf Appendix}

In this appendix we want to give an elementary demonstration that the 
boosts in $SO(4,1)$ correspond to 
time translations for the static observer.
The metric for the static patch is,
\eqn{metric}{ds^2= -(1-r^2)dt^2+\frac{dr^2}{1-r^2}+r^2d\Omega_2^2}
We take $\Lambda/3=1$, where $\Lambda$ is the Cosmological constant. 

The easiest way to think about the de Sitter isometry group ($SO(4,1)$) is 
to think of it as being
embedded in a 5D Minkowski space. In terms of these Minkowski coordinates, the 
static patch can be written as,
\begin{eqnarray}
X^0&=&\sqrt{1-r^2} \sinh t, \\
X^3&=&r \cos \theta, \\
X^1&=&r \sin\theta\cos\phi, \\
X^2&=&r \sin\theta\sin\phi, \\
X^4&=&\sqrt{1-r^2} \cosh t.  
\end{eqnarray}
Its easy to check that $-(X^0)^2+(X^i)^2=1$ and that $-{dX^0}^2+{dX^i}^2$ is equal to
the metric on the static patch. Boosts in $SO(4,1)$ look like,
\eqn{boost}{
\left( 
\begin{array}{c}
{X^0}^{\prime}\\ 
{X^4}^{\prime}
\end{array}
\right)
=
\left(
\begin{array}{cc}
\cosh\beta&\sinh\beta \\ 
\sinh\beta&\cosh\beta
\end{array}
\right)
\left(
\begin{array}{c}
{X^0}\\
{X^4}
\end{array}
\right)
}
Plugging in the expressions for $X^0$ and $X^4$ in terms of $r$ and $t$, multiplying out 
the matrices and simplifying, we end up with,
\eqn{boosted}{
\left(
\begin{array}{c}
{X^0}^{\prime}\\ 
{X^4}^{\prime}
\end{array}
\right)
=
\left(
\begin{array}{c}
\sqrt{1-r^2}\sinh (t+\beta) \\
\sqrt{1-r^2}\cosh (t+\beta)
\end{array}
\right)
,}
which is just the time-translated version of the original expressions.

% ==========================================================================
%

%%%%%%%%%%%%%%%%%%%%%%%%%%%%%%%%%%%%%%%%%%%%%%%%%%%%%%%%%%%%%%%%%%%%%%%%%%%%
%                      REFERENCES                            %
%%%%%%%%%%%%%%%%%%%%%%%%%%%%%%%%%%%%%%%%%%%%%%%%%%%%%%%%%%%%%%%%%%%%%%%%%%%%

\end{document}